# Augmenting Fiat Currency with an Integrated Managed Cryptocurrency

Peter Mell
*National Institute of Standards and Technology*
Gaithersburg MD, USA
peter.mell@nist.gov

*Abstract*—In this work, we investigate how the governance features of a managed currency (e.g., a fiat currency) can be built into a cryptocurrency in order to leverage potential benefits found in the use of blockchain technology and smart contracts. The resulting managed cryptocurrency can increase transparency and integrity, while potentially enabling the emergence of novel monetary instruments. It has similarities to cash in that it enables the general public to immediately transfer funds to a recipient without intermediary systems being involved. However, our system is account-based, unlike circulating bank notes that are self-contained. Our design would allow one to satisfy know your customer laws and be subject to law enforcement actions following legal due process (e.g., account freezing and fund seizure), while mitigating counterparty risk with checks and balances. Funds can thus be transferred only between approved and authenticated users. Our system has on-chain governance capabilities using smart contracts deployed on a dedicated, permissioned blockchain that has different sets of control mechanisms for who can read data, write data, and publish blocks. To enable the governance features, only authorized identity proofed entities can submit transactions. To enable privacy, only the block publishers can read the blockchain; the publishers maintain dedicated nodes that provide access controlled partial visibility of the blockchain data. Being permissioned, we can use a simple consensus protocol with no transaction fees. A separate security layer prevents denial of service and a balance of power mechanism prevents any small group of entities from having undue control. While permissioned, we ensure that no one entity controls the blockchain data or block publishing capability through a voting system with publicly visible election outcomes.

## I. Introduction

Bitcoin is a protocol for a permissionless distributed ledger that was designed to provide non-reversible transactions with direct account-to-account fund transfers where no third party needs to be trusted [1]. It leveraged blockchain technology to enable a form of non-sovereign digital currency that was previously not possible. It and subsequent cryptocurrencies introduced smart contracts and new kinds of decentralized governance models that have significant organizational and political implications (e.g., having no relationship with any government). With respect to these systems, [2] points out that cryptocurrencies can enable users to remain anonymous, can have permissionless access, and thus usually do not support know your customer (KYC) and anti-money laundering (AML) laws at the protocol level by design. In this work, we investigate how to leverage some of the novel benefits provided by blockchain technology and smart contracts to enable a new form of managed cryptocurrency that has built-in support for KYC and AML laws with system governance mechanisms along with a balance of power structure. Note that we are not suggesting that such a cryptocurrency should necessarily be issued, as that decision involves policy and economic factors outside of the scope of this work. Instead, we are proposing a technical architecture that could lead towards the technical ability to do so.

We investigate how the governance features of a managed currency (e.g., a fiat currency) can be built into a cryptocurrency in order to leverage potential benefits found in the use of blockchain technology and smart contracts. It is designed to be compatible with and augment a partner managed currency, the users being able to freely exchange one for the other. The resulting managed cryptocurrency can increase transparency and integrity, while potentially enabling the emergence of novel monetary instruments. It has similarities to cash in that it enables the general public to immediately transfer funds to a recipient without intermediary systems being involved and the associated counterparty risks (a single transaction to the system transfers funds). This is accomplished through a distributed multi-party managed cryptocurrency system providing guarantees similar to Bitcoin style cryptocurrencies. However unlike circulating bank notes, our system is account-based and all recipients are identity proofed and authorized. Our design thus supports the satisfaction of KYC and AML laws at the protocol level. Entities distinct from the platform and currency managers can register as identity providers, ensuring fund transfers only to identity proofed and authenticated recipients while maintaining openness to the private sector and competition. Accounts would also be subject to law enforcement actions following legal due process to include the freezing of accounts and fund seizure.

Our system has on-chain governance capabilities using smart contracts deployed on a dedicated, permissioned blockchain that has different sets of control mechanisms for who can read data, write data, and publish blocks. To enable the cryptocurrency to have built-in governance roles along with KYC/AML checks, only authorized identity proofed entities can submit transactions. To support user privacy features, only the miners (referred to henceforth as validators) can read the blockchain. Validators then maintain dedicated nodes that provide access controlled partial visibility of blockchain data to users (e.g., their account balance, transaction history, and system management transactions). Being permissioned, we can use a lightweight consensus protocol. The protocol could be as simple as the dirty round robin used in Multichain [3]. The use of a security layer that prevents denial of service attacks (which works since all accounts must be pre-authorized and can easily be filtered) can enable a no transaction fee system where the validators are paid by the currency issuer to maintain the currency. Lastly, the architecture contains a balance of power mechanism to prevent any small group of entities from having undue control over the blockchain data or publication of new blocks. While it is a permissioned system, there is not a single entity that decides which accounts can publish blocks. Instead, the existing group of validators vote to determine changes to validator eligibility, with the outcomes being made publicly visible. No one entity controls the blockchain data or block publishing capability.

We implemented our architecture using smart contracts written with the Solidity programming language and made the code open source under a public domain license. It has functions for fund tracking, fiat-to-cryptocurrency fund conversion, transaction logging, account creation, voting scenarios, a bootstrapping mode, role assignment, and the ability of accounts to take special actions given their roles (e.g., law enforcement account freezing and central bank fund creation). A set of initial parameters are used to bootstrap the initial governance options but afterwards a voting system is used for multiple accounts with various roles to collectively manage different aspects of the cryptocurrency.

Research economists seem divided about the effectiveness of central bank issued cryptocurrencies. Some, like in [4], point out the potential economic viability of such assets. Others are more critical: [2] for instance states that there is a 'non-case' for a central bank cryptocurrency; their rationale for this was based on perceived immutable features of cryptocurrencies that would make them useless as alternatives to digital currency. In this work, we want to show that these features, considered immutable, can be altered through changing technical fundamentals about how a cryptocurrency blockchain works; this could enable sovereign cryptocurrencies with fiat currency style governance. Non-sovereign cryptocurrencies started the discussion on how to use blockchain to make the global financial system more stable and distributed; we hope that our work on sovereign cryptocurrencies will further facilitate that discussion.

The remainder of this paper is organized as follows. Section II discusses the foundational technology that underlies the design of our cryptocurrency platform. Section III presents our cryptocurrency architecture while section IV explains the account roles within that architecture. Section V discusses how to instantiate our cryptocurrency to integrate it with a fiat currency. Section VI analyzes the security model of our approach and Section VII discusses our implementation. Section VIII summarizes related efforts and section IX concludes.

## II. FOUNDATIONAL TECHNOLOGY

Our managed cryptocurrency leverages existing approaches and borrows concepts from other technologies. This includes the cryptocurrency role system, on-chain governance of validator nodes, on-chain voting, and the decoupling of the validation and execution of transactions.

Our managed currency relies upon each cryptocurrency account being assigned a set of roles; these roles enable the management and use of the currency. Assigning roles to cryptocurrency accounts was initially introduced in [5].

On-chain governance is used to manage the set of validators through the assignment of 'validator' roles to accounts, those allowed to participate in a consensus algorithm to publish blocks. This concept of on-chain validator governance can be found in the Proof of Authority (PoA) consensus model. Here, block creation is distributed among different allowed nodes over time while offering Byzantine fault tolerance. PoA with smart contract based validator governance is implemented in the Ethereum client Parity (through the Aura consensus algorithm [6]) and by POA Network [7]. Another example of a PoA implementation can be found in the Microsoft Azure Blockchain system [8].

To manage the set of validators as well as other system functions, our managed cryptocurrency smart contracts must implement voting mechanisms which add or remove roles, as well as to approve or disapprove system security actions such as fund transfer reversals. Various standards and projects provide blockchain technology for decentralized voting [9]; the Ethereum standard EIP-1202 [10] offers for example an interface for implementing voting within smart contracts. As another example, the open-source project Aragon [11], built on Ethereum, allows token holders to cast a vote on protocol upgrades by signing a specific transaction. The Delegated Proof of Stake (DPoS) consensus algorithm, used for instance in BitShares [12], is another illustration of an on-chain voting structure. In BitShares, users vote by staking tokens into another account (called 'delegate'); the delegate account is then allowed to execute certain actions on behalf of its stakeholders (such as producing blocks and voting on protocol upgrades).

Lastly, our cryptocurrency introduces the concept of a 'Security Gateway'. A list of gateways linked to their associated validators, maintained at the smart contract level, are charged with pre-processing incoming transactions. This decouples the execution and validation of transactions. The Hyperledger Fabric, a permissioned blockchain, also has a similar decoupling [13]. Note that any mention of commercial products in this paper is for information only; it does not imply recommendation or endorsement.

## III. CRYPTOCURRENCY ARCHITECTURE

A cryptocurrency platform providing the benefits described in Section I and leveraging technology from Section II can be built using the following architecture. Figure 1 shows the overall architecture from the perspective of a single validator.

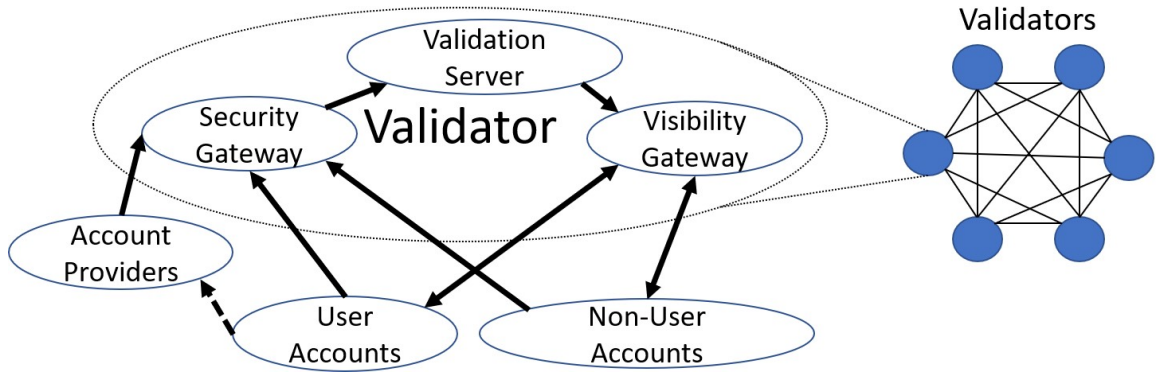

Fig. 1. Interactions of an Individual Validator

### A. Platform Architecture

The architecture requires a permissioned smart contract cryptocurrency platform, such as PoA-based Ethereum. It should be configured to not charge transaction fees or gas for sending transactions to the smart contracts. The native cryptocurrency mechanisms of the platform will not be used and instead the cryptocurrency will be stored within the smart contracts (similar to token based ERC-20 [14] compliant smart contract currencies but running on a dedicated platform). Without gas and transaction fees, validators will be rewarded either off-chain or will participate through being inherently motivated to support the cryptocurrency. This is tractable because our use of a lightweight consensus model makes the execution of a validator node less expensive (it is done this way currently by other permissioned blockchain platforms such as Hyperledger Fabric).

The set of smart contracts will be fixed to a small set used to maintain the cryptocurrency. Being a permissioned system, the block publishing software must determine which validators are allowed to participate in the publication of new blocks. This set of permitted validators is managed at the smart contract layer and can be retrieved from the blockchain. In this way our architecture marries what are usually isolated governance layers, the protocol layer which manages the validator node permissions and the smart contract layer which executes code on behalf of system users.

### B. Security Gateway

Also managed at the smart contract layer is the list of gateways that each validator maintains to accept proposed transactions from the users of the system. These security gateways pre-process incoming transactions to ascertain their validity. Transactions must be properly formatted and are only accepted from accounts that have roles. Security gateways also keep track of the rate of transactions issued by each account. Accounts with an unusually high rate can be throttled as a form of denial of service protection and to prevent any particular account from taking too large a percentage of system resources. It is possible to white list accounts that have a valid reason to issue a high throughput of transactions.

### C. Visibility Gateway

A final platform level resource managed and made visible at the smart contract layer is the set of visibility gateways. Each validator independently maintains a set of such gateways to provide controlled blockchain read access for the account holders. The read access capabilities will be encoded as smart contract view functions (a view function is one that provides read only access and is highly efficient as it is executed locally and not propagated among validators nor included within a block like a normal transaction). However, unlike with typical view function responses, for security reasons (discussed in section VI) the responses will be signed by the associated validator account. The full blockchain is kept private by the validators and user read access is only available through the visibility gateways.

### D. Smart Contracts

The smart contract layer, besides managing the authorized platform level resources listed previously, implements the managed cryptocurrency. The smart contracts maintain the list of authorized accounts, the roles granted to each account with associated features, and the balance in each account. We use the account/balance model as opposed to the unspent transaction output (UTXO) model (e.g., in Bitcoin) to avoid the unnecessary complexity (found in [5]) of having to label each unspent transaction with roles. The roles define a set of permissions that enable certain accounts to manage the cryptocurrency and are discussed in the following section.

### E. Digital Wallets

Identity proofing results in the participants' identifiers being added to a global on-chain registry controlled by the account providers. As described in [15], the identifiers can be held in custodial, semi-custodial, or non-custodial digital wallets that can be integrated into existing applications, browsers, and operating systems.

## IV. MANAGED CRYPTOCURRENCY ROLES

The management features and integration with an associated fiat currency are enabled through accounts with various roles. This account and role capability is instantiated on top of the previously described platform and implemented within the fixed set of smart contracts. Section V will describe how these roles can be used in real world systems.

### A. Platform Managers

An account with the platform manager role sets the policy for the cryptocurrency system and creates accounts and assigns them non-user roles. Policy can be set to be permanent, temporary, or have a timed expiration. Permanent policies cannot be changed once set (assuming the integrity of the blockchain itself is not compromised). They may be used to instantiate a particular architecture that the cryptocurrency will adopt. Alternately, they may be used to provide confidence to the user base that certain features or settings are guaranteed even though the cryptocurrency is managed by a set of privileged entities. Temporary policies can be changed at any time by a currency manager. Timed expiration policies are considered permanent until a published time at which they become temporary. The system may be set up with only one platform manager, a group of accounts that must vote to make changes, or a hierarchical system where higher priority managers can override policies from lower level managers (as in [5]). This latter design can be used as a security feature in case a currency manager account was compromised; higher priority accounts whose keys are stored in physical vaults could be used to override the compromised account and restore the system.

The policies available to be set can include enabling/disabling features within other roles, setting blockchain parameters such as the size and frequency of blocks, adjusting any fees charged (if any), and setting parameters on how voting will be performed (since the system requires groups to vote to perform certain actions).

During the bootstrapping phase for the cryptocurrency, within some fixed number of blocks, the platform manager defines the initial set of validators. Once the bootstrapping phase is over, the accounts with the platform manager role may not modify the validator roles (thus limiting their authority and creating a balance of power).

### B. Account Providers

An account with an account provider role has been authorized by the platform manager(s) to manage user accounts. They identity proof users off-chain, receive a list of the users not yet authorized accounts, and add the user role to those accounts to authorize them. It is important that the users demonstrate ownership of each provided account through proving possession of the associated private key. Each account provider then keeps an internal record of which users are associated with which accounts; this record is not published or shared. This allows KYC and AML laws to be supported at the account provider level (rather than at the entire platform level), which may enhance security and user privacy.

### C. System Security

An account with the system security role has the ability to control other accounts for system security purposes. Such accounts can freeze and unfreeze other accounts. They can also move funds between accounts to confiscate funds or reverse transactions. In the latter case we note that the relevant accounts simply need to be debited and credited funds due to our system being account-based (as opposed to following Bitcoin's UTXO model). We also note that since all accounts are identity proofed, system security actions can take place off-chain using existing legal frameworks.

To limit unauthorized actions, policy can be set by the platform manager requiring an on-chain voting mechanism for certain system security transactions. In addition, the platform manager(s) can limit or disable any of the powers of the system security role through policy settings.

### D. Users

An account with the user role is one that can be used to receive, store, and send value in the form of tokens maintained by the smart contracts. A single user may have multiple accounts and may use multiple account providers to do so (note that every account must be identity proofed by an account provider).

Each account is labelled within the smart contract with its associated public key. A user maintains use of an account through possession of the associated private key (possibly stored on a hardware token for greater security). If a user loses a private key or suspects that their private key has been stolen, they need a way to retake possession of the account. This is accomplished by swapping out the account's original public key with a new one within the smart contract. When creating their account, users can choose what method they prefer to enable this action; there are at least three options. They can trust their account provider to do this for them and simply re-identity proof to their account provider. They could authorize a set of other accounts to validate the public key swap (using accounts they own or accounts of trusted individuals). Or they could require the involvement of system security along with their account provider, necessitating re-identity proofing with both entities.

### E. Currency Managers

An account with the currency manager role has the ability to control the money supply through direct actions or ongoing policy. This includes fund creation, deletion, and the provision of interest. The currency manager accounts vote to set monetary policy or initiate an action (for example the creation of funds to be lent to other entities).

### F. Validators

An account with the validator role is an account that represents an authorized block publisher. Validator accounts vote to add/remove the validator role to/from other accounts. Other than block publishing, the validators manage their respective security and visibility gateways. On their visibility gateways, they make visible all cryptocurrency management transactions to provide full transparency to all users.

Each validator account posts on the smart contract the Internet Protocol (IP) addresses of their security and visibility gateways. The visibility gateways then make the security and visibility gateway addresses visible to all users and the validation server addresses visible to other accounts with the validator role. This latter publication facilitates the peer-to-peer permissioned networking between validators used for transaction propagation and block publication.

Each validator account publishes on the blockchain a publicly visible special public key associated with the signed responses from its visibility gateways. This key is different from the public key for the validator account itself. It also publicly publishes contact information (e.g., an email address) for reporting any problems. This is essential for security reasons discussed in Section VI.

## V. Integration with Fiat Currencies

The architecture presented in sections III and IV is designed to be integrated with a fiat currency and traditional bank deposits. A government administration could instantiate the cryptocurrency and act as the platform manager. The directors of the government's independent central bank could act as the currency managers. The government law enforcement agencies could act in the system security role. This creates a balance of power where no one organization 'controls' the blockchain. To further promote this, government entities separate from the administration can act as the validators (e.g., a set of states). The national standards body can define the specification for the supporting cryptocurrency software and independent laboratories can test compliance of that software. Note that multiple developers should be used, especially for the validator software, for security purposes and the code should be developed open source and made available publicly. This way a single developer cannot maliciously or unintentionally violate the specification and enable non-protocol compliant blocks to be published and accepted.

Financial institutions (e.g., commercial banks, cryptocurrency exchanges, and other fintech companies) could be made account providers, among other entities, since they already must identity proof their customers. They would keep their mapping of identity proofed users to account numbers private and only reveal select information to fulfill a court order (thus supporting KYC and AML laws while still maintaining user privacy). They can modify their banking software to simultaneously show users their bank deposit balances and cryptocurrency balances (since they established each user's accounts). The financial institution itself would not have access to the user's cryptocurrency balance and transactions but their banking application, on behalf of the user, could retrieve this information from the visibility gateways. These applications could then enable the conversion of bank deposits to cryptocurrency and vice versa (also often referred as on-ramp/off-ramp). The application could send cryptocurrency to the financial institution and have the institution deposit traditional money into the user's bank accounts (and vice versa). The application could also transfer funds between the user's different cryptocurrency accounts using a bank owned account as an intermediary to hide any linkage between the user's accounts from appearing on the blockchain. Note that the financial institution obtains cryptocurrency through its existing fiat accounts with the central bank: the institution sends the central bank fiat currency and the central bank sends it cryptocurrency. If users are allowed to interact directly with the central bank, users can perform this operation themselves.

The central bank, as the currency manager, unifies the fiat and cryptocurrency by enabling the exchange between both. The cryptocurrency could be maintained as a separate line item on the central bank balance sheet. The central bank can create and destroy both currencies and thus can implement a cryptocurrency monetary policy in a similar fashion as when managing solely its fiat currency. Note that offering two forms of currency, with different characteristics and risk profiles, can have significant economic implications that are out of scope for this paper.

## VI. Security Analysis

In this section we analyze the security and functionality provided by our architecture using the three traditional computer security pillars of confidentiality, integrity, and availability.

### A. Confidentiality

Our architecture provides accounts/transactions that are pseudonymous for the validators and confidential to the rest of the users. We note that there is a possibility of user confidentiality being lifted when necessary to support KYC and AML laws (e.g., through a court order for validators to reveal transactions and the respective account providers to divulge account ownership). Users choose which account provider they trust to know which accounts they own. The account provider keeps this private unless required to reveal it. Furthermore, the account provider can distribute user funds between the user's accounts such that there is no linkage on the blockchain between the multiple accounts from the same user.

The transactions on the blockchain are kept private and only shared within the set of validators. The visibility gateways only reveal blockchain transactions to the parties involved in those transactions. A downside of this is that it would appear then that accounts with the platform manager, system security, and currency manager roles can issue transactions without oversight. However, the validators are independent entities that make these transactions publicly visible through their visibility gateways. This offers transparency for all management transactions but confidentiality for user fund transfers.

The IP addresses of the validating servers are stored on the blockchain but the visibility gateways make this information visible only to the validator accounts. Thus, the validation servers themselves are kept confidential. If this information was leaked, a single transaction could be used to update a revealed server to a new IP address (to discourage denial of service (DoS) attacks). The security gateways and visibility gateways' addresses are made public. The visibility gateways operate independently with only a copy of the blockchain and thus can be replicated at scale to counter possible DoS attacks.

Likewise, a validator may have multiple security gateways to provide load balancing and DoS protection.

*B. Integrity*

The use of a group of independent validators ensures the integrity of the newly published blocks. No validator will accept a block from another validator that does not follow the established protocol. Each block and the transactions therein must be of the proper form and have the necessary digital signatures. As is the norm with blockchains, each block contains a hash of the previous block to enable detection of any changes with previously published blocks. However unlike public blockchain systems, the users themselves cannot verify the retained integrity of the blockchain and so another mechanism must exist to hold individual validators accountable.

For this, user software will query multiple visibility gateways when retrieving user blockchain data. Any discrepancy between the visibility gateways owned by different validators reveals a problem with one of the validators. An exception is for very recent transactions that have not yet been posted by all validators and thus there should thus be an agreed-upon time delay before taking any action. In the case of a discrepancy, an event transaction is triggered and sent to all validators to describe the discrepancy. As long as at least one validator is honest, the discrepancy will be published. Note that this is considered a 'management' type transaction and thus is made publicly visible to all users. Note that since the visibility gateways sign their responses, the user can prove that multiple visibility gateways provided different answers. As long as a majority of the validators remain honest, the honest ones can vote out any validators that provide incorrect results.

If a set of the validators decide to overtly violate the cryptocurrency protocol (e.g., to change a permanent policy or take control away from the platform or currency managers), this will fork the blockchain as happens with other cryptocurrency systems. The non-violating validators would inform participating parties using off-chain methods and, like other cryptocurrency forks, the resolution would take place off-chain. Given that the cryptocurrency will be tied to a fiat currency, investigations and legal action may be taken against the violating validators. Note that only if 100 percent of the validators collude can they make changes without being noticed. Also, note that our cryptocurrency leverages the fact that it is a sovereign currency, existing within an off-chain legal framework.

*C. Availability*

There are two types of availability that need to be considered: the availability of the cryptocurrency system as a whole and the availability of a particular account to conduct transactions.

The cryptocurrency platform itself has robust availability because it is a distributed system with no central point of failure. Many of the validation servers may fail, even the majority of them, and the system can still continue to function. Note that individual validation servers can be run efficiently due to the use of a lightweight consensus algorithm, permitted by the permissioned blockchain configuration. Also, each security and visibility gateway can be implemented as a cluster of servers to reduce susceptibility to DoS attacks and individual server failures.

Individual accounts are not dependent upon a particular validator and user applications should issue transactions to multiple systems simultaneously (this includes both write transactions to the security gateways and read transactions to the visibility gateways). Using multiple security gateways ensures that no validator could decide to unilaterally block a particular account (note that account ownership is pseudonymous to the validators making this less likely). Using multiple visibility gateways, as discussed above, addresses integrity concerns.

## VII. IMPLEMENTATION

We implemented our managed cryptocurrency as smart contracts using the Solidity programming language. It is available as open source software on Github at https://github.com/usnistgov/managed_token under a public domain license. In this proof-of-concept prototype, we implemented the core functions of the managed cryptocurrency. This includes fund tracking, fiat to cryptocurrency fund conversion, transaction logging, account creation, voting scenarios, bootstrapping mode, role assignment, and the ability of accounts to take special actions given their roles (e.g., law enforcement account freezing and central bank fund creation). Certain aspects are simplified, such as monetary policy options, as our goal was not to create a production system but to demonstrate that this managed architecture approach is feasible. We tested our code by deploying it to a local Ethereum test environment.

A couple of money creation schemes are provided in our system as examples. Schemes are provided to vote and carry out money creation in arbitrary accounts (following a top-down approach) as well as in all user accounts in the form of interests (following a bottom-up approach). Furthermore, these schemes can be either push-based or pull-based. In the push-based model, the money creation function creates funds in the recipient(s) account without any action being required from the recipient(s). In the pull-based model, the currency managers set rules through a single transaction to give the right to the recipient(s) to create their own funds according to this set of rules. This can provide scalability gains as users do not have to claim their allowance right away, and instead, may wait until they need it without any risk of not receiving it. In the case of periodic funds creation (e.g. interests, dividends), a user might be able to skip claiming funds between period X and period X + Y, and then, withdraw funds at period X + Y + 1 for all of the periods between X and X + Y + 1 combined. Finally, a set of view functions allows one to selectively control the visibility of monetary creation and other on-chain fund data, both at the user level and at the currency management level (e.g., global supply indicators).

Note that our implementation did not cover the off-chain aspects of our cryptocurrency architecture. In particular, we did not build the security or visibility gateways (although we did write the smart contract view functions to support the latter). We also did not modify the Ethereum mining software to only publish blocks in collaboration with the validators specified by the smart contracts (but we did implement the smart contract code to enable a set of validators to use the on-chain data to manage themselves).

## VIII. RELATED WORK

Most of the existing work related to managed cryptocurrencies consists in studies and pilots on blockchain-based central bank digital currencies (CBDC), as well as research and development of protocols for stablecoins, algorithmic currency management, and privacy-preserving KYC/AML checks.

The literature distinguishes two main categories of CBDCs: wholesale and retail. As explained by the Bank for International Settlements (BIS) in their money taxonomy [16], a wholesale CBDC is only available to financial institutions and mainly intended for inter-bank transactions whereas a retail CBDC is globally accessible and usable by the general public. Our managed cryptocurrency architecture is geared towards retail CBDCs, although it could also be launched as a wholesale CBDC, at least initially.

We have developed our architecture to change how cryptocurrencies usually work to enable support for retail CBDCs. As stated earlier, this does not mean that we are necessarily claiming that one should be created; we are simply providing some of the technical capability to do so. That said, the subject of state or central bank issued cryptocurrencies has been one of considerable interest. A BIS poll in 2018 showed that more than 70 percent of central banks worldwide were already engaged in CBDC work [17]. Some central banks, such as the central bank of Canada (project Jasper [18]), the Monetary Authority of Singapore (project Ubin [19]), or the Bank of Thailand [20] have focused their research on wholesale CBDC. Also, the European Central Bank and the Bank of Japan are conducting joint research on wholesale CBDCs with Project Stella [21]. Others, such as the Ecuadorian Central Bank ('Dinero Electronico' [22]), the People's Bank of China [23], and the Government of Venezuela (Petro [24]) aim at developing a CBDC for retail use. It should be noted, however, that many central bank efforts do not use (nor plan to use in the future) distributed ledger technologies (DLT); for example the Sveriges Riksbank from Sweden [25] stated that they currently deemed DLT too inefficient for use in a retail CBDC. An example of a non-DLT retail electronic currency is the e-Peso [26]. This is a pilot from the Central Bank of Uruguay that was launched as complement to physical cash but relied on a central registry for ownership recording.

Aside from efforts from governments and central banks, several other blockchain-based research projects have entered the field of managed cryptocurrencies. For example, RScoin [27] provides a cryptocurrency framework using a UTXO model where generation of the monetary supply is controlled by a central authority and transaction processing is handled by dedicated institutions, called 'mintettes'; ultimately, the central authority handles the creation and posting of new blocks. Our system differs from this approach in that currency managers do not influence the block creation process nor benefit from special viewing rights over the content of the blockchain. Another example of a managed cryptocurrency system can be found in Fedcoin [28], which builds upon RScoin's framework by providing a Node.js implementation, KYC rules that enable a central bank to blacklist users, and improved anonymity features. However, unlike our proposal, it does not natively offer the ability for accounts to be assigned roles; this leaves the central bank as the sole entity involved in the identity provider, management, and system security functions (e.g., identity-proofing new accounts, freezing unlawful users, and coin production).

'Decentralized Finance' projects, many of which are currently built with smart contracts deployed on the public Ethereum blockchain or as second layer solutions atop Bitcoin, are also being developed for stablecoins and decentralized currency management where money supply is governed algorithmically (such as Dai [29]). In our system, unlike reserve-backed stablecoins (such as Libra [30], USD Coin [31], and J.P. Morgan Coin [32]) that are pegged one-to-one with the asset(s) that they represent, there is a built-in currency manager role that can develop monetary instruments and vote for monetary policies to increase and decrease the currency supply. Since it is programmable, novel, potentially more flexible monetary instruments may be implemented.

From a security point of view, efforts are being made to offer security standards, toolsets, and services for cryptocurrencies. For example, EIP-1080 [33] is an Ethereum standard that offers an interface geared towards charge back and theft prevention/resolution for ERC-20 tokens [14]. Also, more loosely related is that the Enterprise Ethereum Alliance (EEA) Legal Industry Working Group [34] intends to standardize law-compliant smart contract designs.

Our system provides user privacy through use of a permissioned blockchain that supports roles with different responsibilities and data visibility (e.g., block publishers cannot see account owners' identities). However, cash transactions offer an ideal for anonymity and attempts to achieve this ideal for electronic currencies have been the subject of much research. The development of some privacy-preserving technologies, such as zero-knowledge protocols, has assisted in this objective. For example, Chaum introduced eCash [35] in 1983, one of the first attempts at anonymizing electronic money transactions via the use of blind signatures; Zcash [36] is an example of cryptocurrency that relies on a type of zero-knowledge proof called zk-SNARKs for keeping transactions private; and ChainAnchor [37] offers a method based on the 'Enhanced Privacy ID' zero-knowledge protocol for controlling access to a permissioned blockchain while allowing users to transact pseudonymously and maintain transaction unlinkability.

## IX. Conclusion

Most cryptocurrencies and cryptocurrency research efforts focus on providing cryptocurrencies with strong anonymity and privacy guarantees in a robust distributed system that is not owned or managed by any single entity or group. We do not dispute the importance of such efforts and the emergence of the associated new social constructs, but point out that research in managed cryptocurrencies integrated with our current institutions has been sorely lacking. This is unfortunate as all people live under the laws of their respective countries and it is thus important to research cryptocurrencies that can explicitly support those laws.

In recent years, central banks have been interested in this area, but some of their researchers have discounted cryptocurrency solutions because the foundational technology appears incompatible with central bank goals, especially the support for KYC and AML laws. In this work, we showed how the foundational elements of a cryptocurrency can be rethought to support central bank goals and to explicitly support the laws that apply to electronic fiat currencies. We hope to convince the reader that this type of approach is technically feasible and that cryptocurrencies can be developed that integrate with an associated fiat currency and explicitly support the laws of the respective government.